\input harvmac
\input epsf
\input labeldefs.tmp
\writedefs
\newcount\figno
\figno=0
\def\fig#1#2#3{
\par\begingroup\parindent=0pt\leftskip=1cm\rightskip=1cm\parindent=0pt
\baselineskip=11pt
\global\advance\figno by 1
\midinsert
\epsfxsize=#3
\centerline{\epsfbox{#2}}
\vskip 12pt
{\bf Fig.\ \the\figno: } #1\par
\endinsert\endgroup\par
}
\def\figlabel#1{\xdef#1{\the\figno}}
\def\encadremath#1{\vbox{\hrule\hbox{\vrule\kern8pt\vbox{\kern8pt
\hbox{$\displaystyle #1$}\kern8pt}
\kern8pt\vrule}\hrule}}
\noblackbox
\def\npb#1#2#3{{\it Nucl.\ Phys.} {\bf B#1} (19#2) #3}
\def\plb#1#2#3{{\it Phys.\ Lett.} {\bf B#1} (19#2) #3}

\def\prd#1#2#3{{\it Phys.\ Rev.} {\bf D#1} (19#2) #3}

\def\mpla#1#2#3{{\it Mod.\ Phys.\ Lett.} {\bf A#1} (19#2) #3}

\def\jhep#1#2#3{{\it JHEP} {\bf #1} (19#2) #3}


\def\frac#1#2{{#1 \over #2}}

\def\semi{\subset\kern-1em\times\;}
\def\bar#1{\overline{#1}}

\def\tilde#1{\widetilde#1}

                   \def\CI{{\cal I}}
                   
\def\CK{{\cal K}}

\def\B{{\bf B}}
                     
                     \def\R{{\bf R}}
\def\S{{\bf S}}                     \def\T{{\bf T}}
\def\Z{{\bf Z}}
\def\I{{\rm I}}
\def\K{{\rm K}}
\def\KO{{\rm KO}}
\def\KR{{\rm KR}}
\def\KSp{{\rm KSp}}

\def\O{{\rm O}}
\def\IA{{\rm IA}}
\def\KSC{{\rm KSC}}
\def\TI{{\tilde{{\rm I}}}}
\def\TIA{{\tilde{{\rm IA}}}}
%

\Title{\vbox{\baselineskip12pt
\hbox{hep-th/9902160}
\hbox{CALT-68-2215}}}
{\vbox{\centerline{Brane Transfer Operations and T-Duality}
\bigskip
\centerline{of Non-BPS States}}}
\medskip\bigskip
\centerline{Oren Bergman, Eric G. Gimon and Petr Ho\v rava}
\bigskip
\centerline{\it California Institute of Technology, Pasadena, CA 91125, USA}
\centerline{\tt bergman, egimon, horava@theory.caltech.edu}
\baselineskip18pt
\medskip\bigskip\medskip\bigskip\medskip
\baselineskip16pt
\noindent
Using the relation between D-brane charges and K-theory, we study non-BPS
D-branes and their behavior under T-duality.  We point out that in general 
compactifications, D-brane charges are classified by relative K-theory 
groups.  T-duality is found to act as a symmetry between the relative 
K-theory groups in Type II and Type I/IA theories.  We also study Type 
$\tilde\IA$ theory (which contains an $\O8^-$ plane and an $\O8^+$ plane), 
using K-theory and T-duality to identify its stable D-branes.  Comparison 
with string theory constructions reveals two interesting effects.    
One of them involves the transfer of branes between O-planes, while 
in the other, a D-brane charge which seems conserved near one O-plane in 
fact decays due to the presence of another type of O-plane.  

\Date{February 1999}
\nref\sen{A. Sen, ``Stable Non-BPS States in String Theory,''
\jhep{06}{98}{007}, hep-th/9803194;
``Stable Non-BPS Bound States of BPS D-Branes,'' \jhep{08}{98}{010},
hep-th/9805019;
``Tachyon Condensation on the Brane Antibrane System,''
\jhep{08}{98}{012}, hep-th/9805170;
``$SO(32)$ Spinors of Type I and Other Solitons on Brane-Antibrane Pair,''
\jhep{09}{98}{023}, hep-th/9808141;
``Type I D-Particle and its Interactions,'' \jhep{10}{98}{021},
hep-th/9809111;
``BPS D-Branes on Non-Supersymmetric Cycles,'' hep-th/9812031.}
\nref\ewk{E. Witten, ``$D$-Branes and K-Theory,'' hep-th/9810188.}
\nref\phk{P. Ho\v rava, ``Type IIA D-Branes, K-Theory, and Matrix Theory,''
hep-th/9812135.}
\nref\orennew{O. Bergman and M.R. Gaberdiel, ``Non-BPS States in Heterotic 
-- Type IIA Duality,'' hep-th/9901014.}
\nref\srednicki{M. Srednicki, ``IIB or not IIB,'' \jhep{08}{98}{005}, 
hep-th/9807138.}
\nref\tbls{T. Banks and L. Susskind, ``Brane -- Antibrane Forces,''
hep-th/9511194.}
\nref\greengut{M.B. Green and M. Gutperle, ``Light-Cone Supersymmetry and
D-Branes,'' \npb{476}{96}{28}, hep-th/9604091.}
\nref\gilad{G. Lifschytz, ``Comparing D-Branes to Black-branes,''
\plb{388}{96}{720}, hep-th/9604156.}
\nref\viper{V. Periwal, ``Antibranes and Crossing Symmetry,'' hep-th/9612215.}
\nref\boundictp{E. Gava, K.S. Narain and M.H. Sarmadi, ``On the Bound States
of $p$- and $(p+2)$-Branes,'' \npb{504}{97}{214}, hep-th/9704006.}
\nref\garcia{H. Garc\'\i a-Compe\'an, ``D-Branes in Orbifold Singularities 
and Equivariant K-Theory,'' hep-th/9812226.}
\nref\rmgm{R. Minasian and G. Moore, ``$K$-Theory and Ramond-Ramond Charge,''
\jhep{11}{97}{002}, hep-th/9710230.}
\nref\phopen{P. Ho\v rava, ``Strings on Worldsheet Orbifolds,'' 
\npb{327}{89}{461}; ``Background Duality of Open String Models,'' 
\plb{231}{89}{251}.}
\nref\dlp{J. Dai, R.G. Leigh and J. Polchinski, ``New Connections Between 
String Theories,'' \mpla{4}{89}{2073}.}
\nref\ericcliff{E.G. Gimon and C.V. Johnson, ``K3 Orientifolds,'' 
\npb{477}{96}{715}, hep-th/9604129.}
\nref\kentaro{K. Hori, ``D-Branes, T-Duality, and Index Theory,'' 
hep-th/9902102.}
\nref\ericsh{E.R. Sharpe, ``D-Branes, Derived Categories, and Grothendieck 
Groups,'' \hfill\break
hep-th/9902116.}
\nref\gukov{S. Gukov, ``K-Theory, Reality, and Orientifolds,'' hep-th/9901042.}
\nref\berggab{O. Bergman and M.R. Gaberdiel, ``Stable Non-BPS D-Particles,''
\plb{441}{98}{133}, hep-th/9806155.}
\nref\gp{E.G. Gimon and J. Polchinski, ``Consistency Conditions for 
Orientifolds and D-Manifolds,'' \prd{54}{96}{1667}, hep-th/9601038.}
\nref\karoubi{M. Karoubi, {\it K-Theory. An Introduction\/} (Springer, 1978).}
\nref\husemoller{D. Husemoller, {\it Fibre Bundles\/} (1st ed., Mc~Graw-Hill,
1966; 3rd ed., Springer, 1994).}
\nref\spingeo{H. B. Lawson, Jr.\ and M.-L. Michelsohn, {\it Spin Geometry\/}
(Princeton, 1989).}
\nref\atiyah{M.F. Atiyah, {\it K-Theory\/} (Benjamin, 1964).}
\nref\mfareality{M.F. Atiyah, ``K-Theory and Reality,'' {\it Quart.\ J. Math.\
Oxford} {\bf 17} (1966) 367;\hfill\break
reprinted in \atiyah .}
\nref\ewvect{E. Witten, ``Toroidal Compactification Without Vector
Structure,'' \jhep{02}{98}{006}, hep-th/9712028.}
\nref\psgreen{P.S. Green, ``A Cohomology Theory Based Upon Self-Conjugacies
of Complex Vector Bundles,'' {\it Bull.\ Amer.\ Math.\ Soc.} {\bf 70}
(1964) 522.}
\nref\anderson{D.W. Anderson, ``The Real K-Theory of Classifying Spaces,''
{\it Proc.\ Nat.\ Acad.\ Sci.} {\bf 51} (1964) 634.}
\nref\piljin{P. Yi, ``Membranes from Five-Branes and Fundamental Strings from 
D$p$-Branes,'' hep-th/9901159.}
\newsec{Introduction}

\noindent 
A closer look at the dynamics of various unstable D-brane systems (such as
brane-antibrane systems) has recently opened a new perspective for
understanding D-branes and their conserved charges.  Traditionally,
D-branes are understood as RR-charged stringy solitons on which strings can
end;  in the new framework \refs{\sen,\ewk,\phk,\orennew}, D-branes appear as
topological defects in the worldvolume of unstable brane systems of higher
dimension.

A crucial role in this construction is played by our improved understanding of
the string theory tachyon \sen\ (see also \refs{\srednicki-\boundictp}).  
String theory has been plagued with tachyons since its early days, but whether 
they represent an incurable instability of the
theory or have a more subtle role in the dynamics was not known.  It
is now believed that the tachyonic mode of the open string stretching between
a D-brane and a D-antibrane (or between a pair of unstable D-branes) is 
really a legitimate Higgs field, and therefore does not represent an incurable 
instability.  Instead, it tends to develop a stable vacuum expectation value, 
leading to the decay of the unstable state into a stable state.  Depending on 
the details of the original unstable configuration, the resulting stable state 
can contain topological defects that correspond to stable D-branes.  

Any such construction can be related through a hierarchy of embeddings to
bound states in the unstable system of a number of spacetime-filling
D9-branes.  The worldvolume dynamics of this system contains 
$U(N)\times U(N)$ Yang-Mills theory and a Higgs field (a.k.a.\ ``tachyon'') 
in the $({\bf N},\bar {\bf N})$ representation in the case of Type IIB
theory \refs{\sen\ewk}, and $U(N)$ Yang-Mills theory with an adjoint
Higgs in the Type IIA case \phk .  All possible stable D-branes -- both
supersymmetric {\it and\/} non-supersymmetric -- appear as topological defects
in the worldvolume Higgs field on these spacetime-filling D-brane systems.  In
this sense, the spacetime-filling brane system provides a universal medium in
which all stable D-brane charges are carried by conventional topological
defects, similar to vortices in Type II superconductors or magnetic monopoles
of grand unified theories.

The precise dynamics of these unstable D-brane systems is not known, but 
the topological information needed for the complete classification of D-brane 
charges can still be determined.  This information is usefully encoded in 
K-theory \refs{\ewk,\phk,\garcia}. (The connection between D-brane charges and
K-theory was first suggested \rmgm .) Once one identifies the 
K-theory group relevant to a given compactification, one can use methods 
developed in the mathematical literature to compute it, and thereby determine 
the spectrum of conserved D-brane charges.  Having classified the charges, one 
can then look for a string theory construction of the corresponding D-branes.  

One of the defining qualities of D-branes, which in fact is how they were 
discovered \refs{\phopen,\dlp}, is their transformation under T-duality.  
Because T-duality exchanges Neumann open string boundary conditions with 
Dirichlet ones, it exchanges wrapped branes and unwrapped branes.  One of the 
purposes of this paper is to determine how T-duality is manifested in the 
bound state construction above.  
We shall therefore study compactifications on $X\times\S^1$, and orientifolds 
thereof.  

In Section~2 we discuss the K-theory realization of T-duality in Type II 
string theory.  As one of the central points of the paper, we show that 
in general string theory compactifications, D-brane charges are classified 
by {\it relative\/} K-theory groups, 
such as $\K(\S^p\times Y,\,Y)$, where $Y$ is the compactification manifold.  
In the case of Type II strings compactified on a circle, the relative K-theory 
group of D-brane charges splits into the sum of two groups, whose 
elements reflect the split of the D-brane charges between wrapped D-branes
and unwrapped D-branes.  The K-group for IIB on a circle,
$\K(X\times\S^1,\,\S^1)$, is isomorphic to the K-group for IIA on a circle,
$\K^{-1}(X\times\S^1,\,\S^1)$, clearly in line with T-duality.  Moreover, the
isomorphism exchanges the subgroup associated with wrapped D-branes
on one side with the subgroup for unwrapped D-branes on the other side.  In 
the rest of this paper, we would like to see if this clear split between 
wrapped and unwrapped D-branes holds generically in K-theory.

Our first test case, discussed in Section~3, involves looking at the T-duality 
between Type I strings on a circle and the Type IA orientifold with two 
$\O8^-$ orientifold planes.  Unlike in the Type II theory, there are now new 
non-BPS D-branes with $\Z_2$ valued charges.  At first, the situation looks 
quite similar to the Type II case: The K-group for Type I on a circle,
$\KO(X\times \S^1,\,\S^1)$, is again found to be isomorphic to the
Type IA K-group $\KR^{-1}(X\times\S^1,\,\S^1)$, and it splits into 
two parts which one may naively interpret as corresponding to
wrapped branes and unwrapped branes.

This interpretation raises two puzzles.  First, some of the non-BPS D-branes 
that were stable in flat space are now only 
stable for a certain range of the circle's radius, even though they carry a 
conserved charge.  Second, in Type IA we expect to find unwrapped branes 
localized on each of the two orientifold planes, yet the corresponding charges 
seem to be missing from the subgroup we naively associate with unwrapped 
branes.  Section~3 will demonstrate how these two puzzles are resolved.  A 
key element in this resolution involves processes that we refer to as brane 
transfer operations.  

Our second test case, presented in Section~4, deals with the more
exotic Type $\tilde\I$ open string theory, which is T-dual to the Type
$\tilde\IA$ orientifold with both an $\O8^-$ and an $\O8^+$
orientifold plane. The K-group associated with these theories,
which turns out to be equivalent to a certain K-theory group known
in the mathematical literature as $\tilde\KSC(X)$, does not split
naturally into the of sum of two sub-groups.  The split between
wrapped branes and unwrapped branes is totally defeated. We will
show that the root of this problem is linked to the fact that some
non-BPS brane configurations locally stable at the $\O8^-$
orientifold plane become unstable due to the presence of the
$\O8^+$ plane and vice versa.  This phenomenon can have important 
consequences for the piecewise analysis of the stable non-BPS D-brane spectra 
in various compactifications, such as when one approximates singularities in 
K3 orientifolds by ALE spaces \ericcliff .

Many technical details required for our analysis have been relegated to 
an appendix, which also serves as a collection of basic facts in K-theory, 
and can therefore be of some independent interest.  

While this paper was being written, two papers \refs{\kentaro,\ericsh} 
appeared in which some overlapping results on T-duality in K-theory were 
obtained.  The connection between KR-theory and orientifolds has also been 
discussed in \gukov.   

\newsec{Type II Theories}

\noindent As a warm-up exercise, we set the stage for our later analysis of 
stable D-branes in various orientifold models by first analyzing the case of 
Type II theories compactified on a circle to nine dimensions.  (As we will 
see, this procedure can be easily iterated to understand $\T^n$ 
compactifications).   It turns out that all stable D-brane states predicted 
by K-theory in these compactifications carry conventional Ramond-Ramond 
charges.  Therefore, we do not expect any surprises; the main goal in 
this brief section is to see that T-duality of Type II theories is indeed a 
manifest symmetry in K-theory.  

\subsec{$D={}$10}

\noindent Stable D$p$-brane charges in Type IIA and Type IIB theory on
$\R^{10}$ are classified by the K-theory groups $\K^{-1}(\S^{9-p})$ and 
$\tilde\K(\S^{9-p})$ 
respectively, where the sphere $\S^{9-p}$ represents the dimensions
transverse to the worldvolume of the $p$-brane, compactified by adding a
point at infinity.  This result can be derived by realizing supersymmetric 
Type II D-branes as stable topological defects in the Higgs field on the 
worldvolume of a system of spacetime-filling D9-branes \refs{\ewk,\phk}.
Central to this derivation is the relation (reviewed in the appendix)
between homotopy theory, which classifies topological defects, and K-theory,  
which classifies configurations of spacetime-filling branes up to creation and 
annihilation.  According to this relation, the K-theory groups of spheres are 
equal to the homotopy groups of the vacuum manifold of the Higgs field that 
appears in the worldvolume of the corresponding system of spacetime-filling 
branes.  In Type IIB theory \ewk , the Higgs field is in the $({\bf N},\bar 
{\bf N})$ of the $U(N)\times U(N)$ gauge group, and its vacuum manifold is a 
copy of $U(N)$.  Its homotopy groups are related to K-theory via 
\eqn\khomotopy{\tilde{\K}(\S^n) = \pi_{n-1}(U(N)).}
In Type IIA theory \phk , the Higgs field is in the adjoint of the $U(2N)$ 
gauge group, and the vacuum manifold is given by the group coset 
$U(2N)/U(N)\times U(N)$.  This is in turn related to K-theory groups by 
\eqn\konehomotopy{\K^{-1}(\S^n) = \pi_{n-1}\left(U(2N)/U(N)\times U(N)\right).}

These K-theory groups, shown in Table~\eetabone , reproduce the known 
spectrum of BPS D$p$-branes in the Type II theories in $\R^{10}$, with 
$p=0,2,4,6,8$ in Type IIA and $p=-1,1,3,5,7$ in Type IIB.%
\foot{We have included the charge  of the spacetime-filling Type IIA D9-branes 
in Table~\eetabone\ (and will do so systematically in similar 
cases throughout the paper), even though the absence of a net 
9-brane charge in Type IIB theory is forced by the condition of
tadpole anomaly cancellation.}
\eqn\eetabone{\vbox{\offinterlineskip \hrule
\halign{&\vrule#&\strut\ \ \hfil#\ \ \cr
height2pt&\omit&&\omit&&\omit&&\omit&&\omit&&\omit&&\omit&&\omit&&\omit&
&\omit&&\omit&&\omit&\cr &D$p$-brane\qquad
&&D9&&D8&&D7&&D6&&D5&&D4&&D3&&D2&&D1&&D0&&D$(-1)$&\cr
\noalign{\hrule}
height2pt&\omit&&\omit&&\omit&&\omit&&\omit&&\omit&&\omit&&\omit&&\omit&
&\omit&&\omit&&\omit&\cr & Transverse {\bf X}
&&$\S^0$&&$\S^1$&&$\S^2$&&$\S^3$&&$\S^4$&&$\S^5$&&$\S^6$&
&$\S^7$&&$\S^8$&&$\S^9$&&$\S^{10}\ $&\cr \noalign{\hrule}
height2pt&\omit&&\omit&&\omit&&\omit&&\omit&&\omit&
&\omit&&\omit&&\omit&&\omit&&\omit&&\omit&\cr &$\tilde\K({\rm\bf
X})$\qquad\quad\ \ &&\Z\ &&0 &&\Z\ &&0 &&\Z\ &&0 &&\Z\ &&0 &&\Z\
&&0 &&\Z\ \ \ &\cr \noalign{\hrule}
height2pt&\omit&&\omit&&\omit&&\omit&&\omit&&\omit&
&\omit&&\omit&&\omit&&\omit&&\omit&&\omit&\cr &$\K^{-1}({\rm\bf
X})$\qquad\ \ &&0 &&\Z\ &&0 &&\Z\ &&0 &&\Z\ &&0 &&\Z\ &&0 &&\Z\
&&0\ \ \ &\cr} \hrule}}

\subsec{General Compactifications}

\noindent The next step would be to consider Type II theory compactified on 
$\S^1$.  Before we discuss this case in detail, it seems worthwhile to first 
study the classification of D-brane charges for a more general compactification
space $Y$ of dimension $d$.
We are interested in finding all D-brane charges of codimension $n$ 
in the non-compact space $\R^{9-d}$.  
Such charges will arise both from D-branes located at particular points 
in $Y$, and from D-branes which wrap non-trivial cycles in $Y$.
Since we are only interested in objects of finite energy (or action), 
we consider only configurations that are equivalent to the vacuum 
asymptotically in the transverse space $\R^n$,
i.e., along a copy of the entire compactification manifold $Y$ at infinity.  
Therefore $\R^n$ is effectively replaced with $\S^n$ by adding a point at 
infinity, which corresponds in the full theory to 
adding a copy of the compactification manifold $Y$ at infinity. 
In mathematical terms, this 
requires us to consider bundles which are trivialized on the compactification 
manifold $Y$ at infinity; such bundles define groups known in the mathematical 
literature as {\it relative\/} K-theory groups (cf.\ the appendix).  Thus, we 
conclude that the proper way of understanding the spectrum of D-brane charges 
is in terms of relative K-theory.  In Type IIB and Type IIA theory on $Y$, the 
relative groups that classify D-brane charges are denoted by 
$\K(\S^n\times Y,\,Y)$ and $\K^{-1}(\S^n\times Y,\,Y)$, respectively.

The argument leading to the appearance of relative K-theory groups is 
essentially independent of the type of string theory considered.  It suggests 
the following prescription for 
identifying stable D-brane charges in general string theory compactifications
on $\R^{9-d}\times Y$, at least when no RR backgrounds or non-trivial $B_{\mu\nu}$
backgrounds are excited: {\it Stable charges carried by D-branes of
codimension $n$ in the non-compact dimensions are classified by the relative
K-theory groups $\CK^{-q}(\S^n\times Y,\,Y)$.}  The value of $q$ and the type
$\CK$ of K-theory depends on the type of string theory and the compactification
manifold $Y$.

\subsec{$D={}$9}

\noindent
Having clarified the appearance of relative K-theory groups in the 
classification of D-brane charges in general compactifications, we can now 
return to Type II theory on a circle.  Using \eeredprodk\ and arguments 
presented in the appendix, the relative groups that classify D-brane 
charges in Type IIB and Type IIA theory on $\S^1$ can be shown to decompose 
as follows:
\eqn\eekunnk{\K(X\times\S^1,\S^1)=\K^{-1}(X)\oplus\tilde\K(X),}
and
\eqn\eekunnkone{\K^{-1}(X\times\S^1,\S^1)=\tilde\K^{-2}(X)\oplus\K^{-1}(X),}
where in both cases the first term is the contribution of unwrapped branes
to the nine-dimensional D-brane charge,
and the second term is the contribution of wrapped branes.
Since by Bott periodicity $\tilde\K^{-2}(X)\cong\tilde\K(X)$,
the above groups are isomorphic
\eqn\eetdualityii{\K(X\times\S^1,\S^1)\cong\K^{-1}(X\times\S^1,\S^1),}
and we recover the result that the spectrum of D-brane charges in nine
dimensions is identical for Type IIA and Type IIB.  In fact, for each 
$X=\S^n$, the relative K-theory group of D-brane charges is $\Z$  -- the RR 
charge of the corresponding D$p$-brane.
  
Furthermore, since the above isomorphism maps the first (second)
term in \eekunnk\
to the second (first) term in \eekunnkone , and therefore exchanges
unwrapped and wrapped D-branes, it corresponds precisely to T-duality.
More rigorously, this follows from a derivation of \eetdualityii\ that keeps 
track of the multiplicative structure of K-theory (see \eemultipk\ of the 
appendix and the discussion therein).  

\subsec{$D<{}$9}

\noindent We can iterate the steps of the previous subsection, and extend our
results to higher toroidal compactifications.  Thus, the relative group of
D-brane charges in Type IIB theory is
\eqn\eetork{\K(X\times\T^m,\,\T^m)=\bigoplus_{n=0}^m\pmatrix{m\cr n\cr}\tilde
\K^{-n}(X)=
\tilde\K(X)^{\oplus 2^{m-1}}\oplus \K^{-1}(X)^{\oplus 2^{m-1}},}
where the second equality follows by Bott periodicity.  An analogous 
calculation on the Type IIA side (cf.\ the appendix) reveals
\eqn\eetorka{\K{}^{-1}(X\times\T^m,\,\T^m)\cong\K(X\times\T^m,\,\T^m).}
This proves T-duality of D-brane charges in Type II theory on $\T^m$, and
gives the expected degeneracy of D$p$-brane charges arising from wrapping all
higher supersymmetric branes on various cycles of the torus.  All in all,
this shows that in the case of Type II theory on $\T^m$, D-brane 
charges predicted by the more 
precise K-theory arguments coincide with those predicted by the somewhat
cruder argument that relates D-brane charges to RR charges (and therefore to
the cohomology of the compactification manifold).

\newsec{Type I Theory and its T-Duals}

\noindent
Our next case study is Type I string theory. Here we encounter two
new features: $\Z_2$-charged non-BPS D-branes, and discrete
$\Z_2$-valued Wilson lines on some of the wrapped D-branes. The
latter would seem to require additional $\Z_2$ charges in the
D-brane spectrum, naively absent in K-theory. Surprisingly, we
shall see that the $\Z_2$ charges of unwrapped non-BPS D-branes
already incorporate the $\Z_2$ Wilson lines of wrapped D-branes.
In T-dual orientifolds this is seen via ``brane transfer
operations,'' whereby an unwrapped brane at one orientifold plane
is ``transferred'' by a wrapped brane to another orientifold
plane.

\subsec{D=10}

\noindent
In Type I theory, the full spectrum of D-brane charges can be determined from 
the dynamics of unstable systems of multiple D9-brane D$\bar 9$-brane pairs.  
Since the action of the orientifold group is antilinear on Chan-Paton bundles, 
the K-theory that arises in such systems is the KO-theory of real virtual 
bundles \ewk.  The KO-groups of spheres take the following values, 
\smallskip
\eqn\eetabtwo{\vbox{\offinterlineskip \hrule
\halign{&\vrule#&\strut\ \ \hfil#\ \ \cr
height2pt&\omit&&\omit&&\omit&&\omit&&\omit&&\omit&&\omit&&\omit&&\omit&
&\omit&&\omit&&\omit&\cr &Dp-brane\qquad&&D9&&D8&&D7&&D6&&D5&&D4
&&D3&&D2&&D1&&D0&&D$(-1)$&\cr \noalign{\hrule}
height2pt&\omit&&\omit&&\omit&&\omit&&\omit&&\omit&&\omit&&\omit&&\omit&
&\omit&&\omit&&\omit&\cr &Transverse {\bf X}
&&$\S^0$&&$\S^1$&&$\S^2$&&$\S^3$&&$\S^4$&&$\S^5$&
&$\S^6$&&$\S^7$&&$\S^8$&&$\S^9$&&$\S^{10}\ $&\cr
\noalign{\hrule}
height2pt&\omit&&\omit&&\omit&&\omit&&\omit&&\omit&
&\omit&&\omit&&\omit&&\omit&&\omit&&\omit&\cr &$\tilde\KO({\rm\bf
X})$\qquad\ \ &&\Z\ &&$\Z_2$&&$\Z_2$&&0\ &&\Z\ &&0\ &&0\ &&0\
&&\Z\ &&$\Z_2$&&$\Z_2\ \ $&\cr
}\hrule}}
The values for $\tilde\KO(\S^n)$ reproduce the known BPS
D$p$-branes of ten-dimensional Type I string theory, namely those
with $p=1,5,9$. In addition, we encounter $\Z_2$-charged non-BPS
D$p$-branes with $p=-1,0,7$ and $8$. The former carry RR charge and
correspond to boundary states of the form
\eqn\bps{|Dp\rangle = {1\over\sqrt{2}}(|Bp\rangle_{NSNS} \pm
 |Bp\rangle_{RR}),}
where the relative sign differentiates brane from antibrane,
whereas the latter do not carry RR charge, and therefore correspond to
states of the form
\eqn\nonbps{|Dp\rangle=|Bp\rangle_{NSNS},}
and are their own antibranes.  
All the properties of the non-BPS D-branes can be obtained from
these boundary states via tree-level overlaps with other boundary
states \refs{\berggab,\orennew,\sen} (for an extension to higher loops, 
see \ewk ).  This construction proves that all charges from Table 
\eetabtwo\ are carried by D-branes, i.e., spacetime defects on which 
strings can end.  

The most useful description of the non-BPS D-branes is often in terms of bound 
states of a {\it single\/} BPS D-brane D-antibrane pair with lowest possible 
dimension.  In this approach, the D0-brane and D8-brane simply 
correspond to topologically stable kinks in the tachyon 
(Higgs) field living on the worldvolume of the D1-D$\bar 1$ and D9-D$\bar 9$ 
systems, respectively \sen . The 
D$(-1)$-brane and D7-brane, on the other hand, correspond to the
D$(-1)$-D$\bar{(-1)}$ and D7-D$\bar 7$ systems in 
Type IIB, respectively, projected by $\Omega$ \ewk .  This approach allows 
one to easily deduce the worldvolume theories of the 
non-BPS D-branes, and in particular the worldvolume gauge groups:
\eqn\gaugegroups{\vbox{\offinterlineskip \hrule
\halign{&\vrule#&\strut\quad\hfil#\quad\cr
height2pt&\omit&&\omit&&\omit&&\omit&&\omit&&\omit&&\omit&\cr
&Dp-brane\ &&D0&&D1&&D5\quad&&D7\ &&D8&&D9&\cr \noalign{\hrule}
height2pt&\omit&&\omit&&\omit&&\omit&&\omit&&\omit& &\omit&\cr
&Gauge group&&$\Z_2$&&$\Z_2$&&$USp(2)$&&$U(1)$&&$\Z_2$&&$\Z_2$&\cr
}\hrule}}

\subsec{$D={}$9}

\noindent
As in the Type II case, D-brane charges of Type I compactified on
a circle are classified by the relative K-theory group
$\KO(X\times\S^1,\,\S^1)$. This group is evaluated in the
appendix, giving:
\eqn\eekunnko{\KO(X\times\S^1,\S^1)=\tilde\KO{}^{-1}(X)\oplus\tilde{\KO}(X).}
Thus, we obtain the following nine-dimensional stable D-brane
charge spectrum:
\eqn\eetabtwohalf{\vbox{\offinterlineskip \hrule
\halign{&\vrule#&\strut\ \ \hfil#\ \ \cr
height2pt&\omit&&\omit&&\omit&&\omit&&\omit&&\omit&&\omit&&\omit&&\omit&
&\omit&&\omit&\cr &Dp-brane\ \
&&D8&&D7&&D6&&D5&&D4&&D3&&D2&&D1&&D0&&D$(-1)$&\cr
\noalign{\hrule}
height2pt&\omit&&\omit&&\omit&&\omit&&\omit&&\omit&&\omit&&\omit&&\omit&
&\omit&&\omit&\cr &Transverse {\bf X}
&&$\S^0$&&$\S^1$&&$\S^2$&&$\S^3$&&$\S^4$&&$\S^5$&&$\S^6$&&$\S^7$&
&$\S^8$&&$\S^9\ \ $&\cr
\noalign{\hrule}
height2pt&\omit&&\omit&&\omit&&\omit&&\omit&&\omit&
&\omit&&\omit&&\omit&&\omit&&\omit&\cr &$\tilde\KO({\rm\bf X}
)$\quad &&\Z\ &&$\Z_2$&&$\Z_2$&&0 &&\Z\ &&0 &&0 &&0 &&\Z\
&&$\Z_2\ \ $&\cr \noalign{\hrule}
height2pt&\omit&&\omit&&\omit&&\omit&&\omit&&\omit&
&\omit&&\omit&&\omit&&\omit&&\omit&\cr &$\tilde\KO^{-1}({\rm\bf X}
)$\quad &&$\Z_2$&&$\Z_2$&&$0$ &&$\Z$ &&0 &&0 &&0 &&\Z\
&&$\Z_2$&&$\Z_2\ \ $&\cr }\hrule}}
Note that the relative K-theory groups correctly include the nine-dimensional 
D$p$-brane charges that correspond to unwrapped D$p$-branes as well as wrapped 
D$(p+1)$-branes in the ten-dimensional theory.  

Under T-duality, Type I string theory is mapped to an orientifold
of Type IIA of the form $\R^{9}\times \S^1/\Omega\cdot\CI$, known as Type IA
(or Type I${}'$); here $\Omega$ acts as a reflection on the
worldsheet, and $\CI$ acts as a reflection on the compact
direction. The compact direction is therefore an interval, rather
than a circle.  The associated relative K-theory group is given by
$\KR^{-1}(X\times\S^{1,1},\,\S^{1,1})$ \refs{\ewk,\phk}.%
\foot{Here we use standard mathematical parlance \mfareality\ to denote by 
$\S^{1,1}$ the unit circle inside the plane $\R^{1,1}$ where the $\KR$ 
involution leaves the first coordinate invariant and reflects the second 
(cf.\ the appendix).}
(This follows directly from the action of the orientifold group on the system 
of unstable D9-branes of Type IIA theory, as was briefly pointed out in \phk .)
We show in the appendix that the Type IA K-group decomposes as follows: 
\eqn\eekunnkr{\KR^{-1}(X\times\S^{1,1},\,\S^{1,1})=\tilde{\KO(X)}
  \oplus\tilde\KO{}^{-1}(X).}
Therefore, as in the Type II case, T-duality between Type I and Type IA theory
manifests itself as an isomorphism between the relative K-theory groups,
\eqn\eeisomori{\KO(X\times\S^1,\,\S^1)\cong\KR^{-1}(X\times\S^{1,1},\,
\S^{1,1}),}
whose elements correspond to D-brane charges of the orientifold 
compactification.  

The isomorphism \eeisomori\ again maps the first term of the relative KO-group
in \eekunnko\ to the second term of the relative KR-group in
\eekunnkr , and vice versa. It is therefore tempting to identify
the respective terms as the contributions to nine-dimensional
D-brane charges coming from unwrapped and wrapped ten-dimensional
D-branes. For example, nine-dimensional 0-brane charge in Type I
receives a $\Z_2$ contribution from the unwrapped non-BPS
D0-brane, and a $\Z$ contribution from wrapped BPS D1-branes. In
the Type IA description, it receives contributions from the T-dual
configurations, i.e., $\Z_2$ from the wrapped non-BPS D1-brane, 
and $\Z$ from unwrapped BPS D0-branes. However, on the face of it, there 
seems to be a problem with this interpretation: the non-BPS D-branes are not 
stable for all radii, and therefore cannot contribute conserved charges
everywhere in moduli space. What is then responsible for the
$\Z_2$ charges when the non-BPS D-branes are unstable? To answer
this question, let us first recall how non-BPS D-branes decay.

\subsec{D-brane decay}

\noindent
Consider for example the non-BPS D0-brane in Type I. The spectrum
of open strings beginning and ending on the D0-brane is
tachyon-free in ten dimensions. Once we compactify on a circle
however, the ground state at winding number 1 will have a classical mass 
squared given by ($\alpha'=1$)
\eqn\eetachyon{m^2 =-\frac{1}{2}+R^2,}
and will therefore become tachyonic when $R<1/\sqrt{2}$.  As a result, the
D0-brane should then decay into a wrapped D1-D$\bar 1$ system.
Recall, however, that in describing the D0-brane as a
D1-D$\bar 1$ bound state one requires the tachyon (Higgs)
field of the D1-D$\bar 1$ system to condense into a kink.
This implies anti-periodic boundary conditions for the tachyon,
achieved by turning on a $\Z_2$ Wilson line on either the D1-brane
or the D$\bar 1$-brane.  The $\Z_2$ charge of the D0-brane
is not lost when it decays, but rather reappears as a
$\Z_2$-valued Wilson line on its decay products. The unwrapped
D8-brane and D7-brane meet a similar fate when $R<1/\sqrt{2}$;
they decay into a wrapped D9-D$\bar 9$ and D8-D8 system,
respectively, also with a non-trivial $\Z_2$ Wilson line.

\fig{Decay of non-BPS D-branes in Type I and Type
IA}{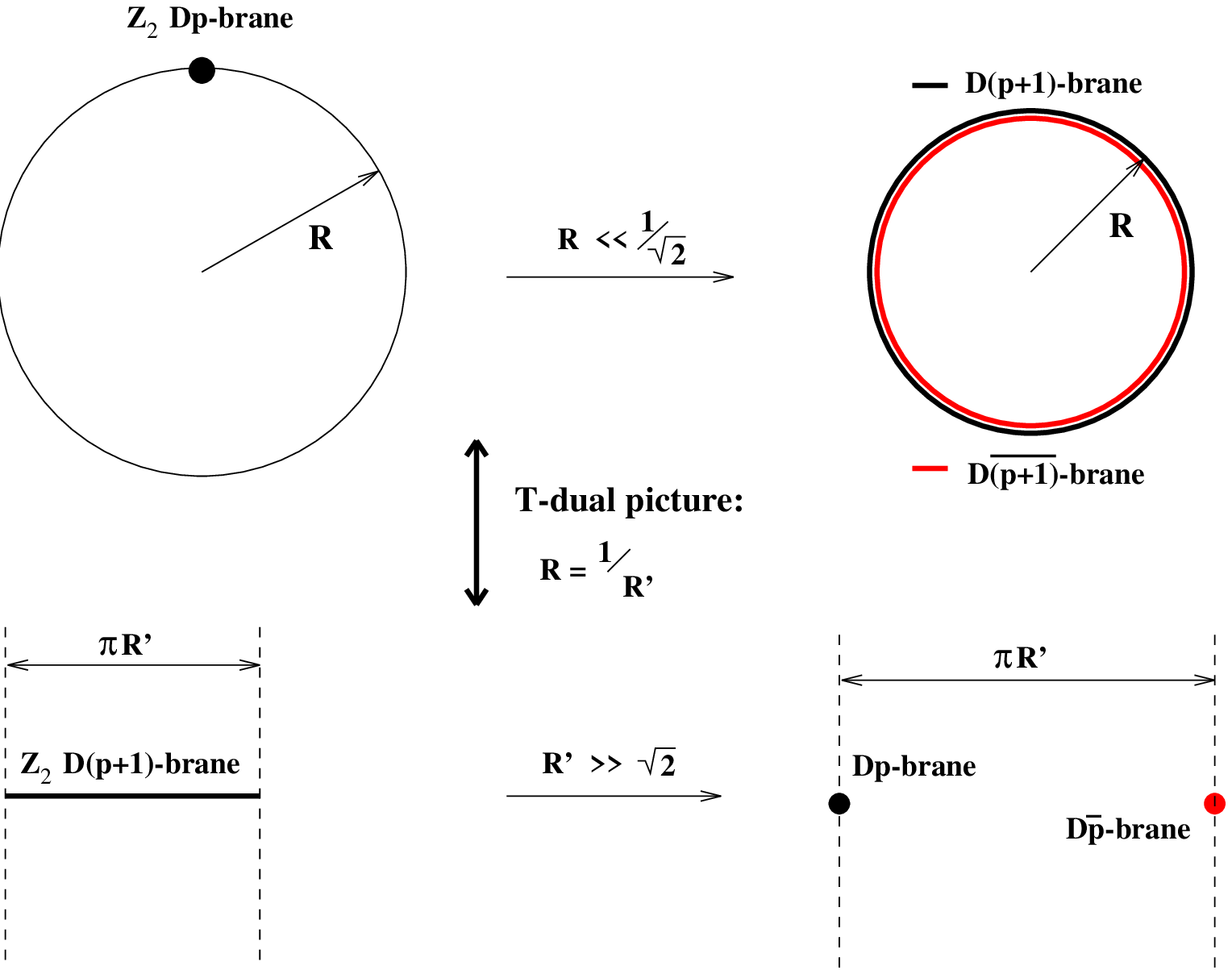}{4.5truein}

Conversely, after T-duality the above unwrapped non-BPS D$p$-branes
become wrapped non-BPS D$(p+1)$-branes of Type IA with $p+1=1,8,9$.
These develop a tachyon of unit momentum when the dual radius
becomes too large ($R'>\sqrt{2}$), and consequently decay into an
unwrapped D$p$-D$\bar p$ system restricted to the 
orientifold planes. The non-trivial $\Z_2$ Wilson line in the Type
I description reflects the presence of the resulting D-brane and
D-antibrane on {\it different} orientifold planes. We conclude that in
the Type IA picture the $\Z_2$ charge of the decaying D-brane is
encoded in the $\Z_2$ choice of locations for its decay products.
These decay processes are summarized in Fig.~1.

For the Euclidean wrapped $\Z_2$ D0-brane and the $\Z_2$ D-instanton,
compactification on a circle again introduces regions of ``stability'' on
the moduli space of the circle, but since these are instantonic 
configurations, we now compare the values of the instanton action.

\subsec{D-brane transfer}

\noindent
The picture of D-brane decay described above offers insight for
the resolution of another puzzle which is most clearly 
illustrated in the Type IA picture.  When $R'<\sqrt{2}$, there seem to be two 
distinct sources of $\Z_2$ charge associated with D0-branes in the 
nine-dimensional theory.  The first one is due to the possibility of locating 
a single D0-brane at either orientifold plane, while the second is due to the 
stretched non-BPS D1-brane.  However, K-theory indicates that there is only 
one D0-brane $\Z_2$ charge.  

As we saw above, when $R'>\sqrt{2}$ the D1-brane decays 
into a D0-brane at one O8-plane, and a D$\bar 0$-brane at the other O8-plane.  
This is a crucial clue for the resolution of our puzzle.  
Consider a configuration consisting of a stuck D0-brane (half D0-brane) at
one orientifold plane and a wrapped non-BPS D1-brane. As far as conserved 
D-brane charges are concerned, this configuration is 
completely equivalent to a stuck D0-brane at the other orientifold
plane, and in fact unstable to decay into it. The same is true for
the unwrapped D7 and D8-brane. In each case, a D-brane stuck at
one orientifold plane is ``transferred'' by a wrapped D-brane of
one higher dimension to the other orientifold plane. Thus we see
that the puzzle of missing $\Z_2$ charges in K-theory is resolved by a ``brane
transfer operation'' (Fig.~2). \fig{D-brane transfer operation in
Type IA}{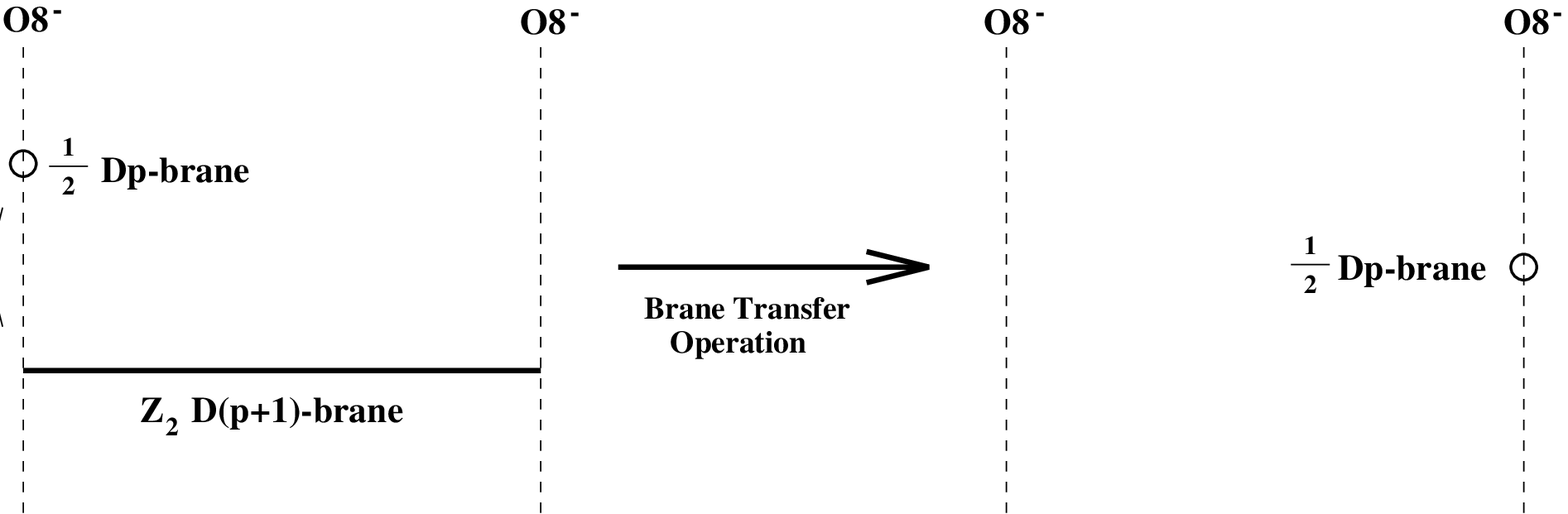}{4.5truein}

\subsec{$D<{}$9}

\noindent
Our analysis can be straightforwardly extended to T-duality of Type I 
theory on higher tori.  D-brane charges in Type I theory on $\T^m$ correspond 
to the relative K-theory group (calculated by iterating \eeredprodko )
\eqn\eetorko{\KO(X\times\T^m,\,\T^m)=\bigoplus_{p=0}^m\pmatrix{m\cr p\cr}
\tilde\KO{}^{-p}(X),}
to be compared to the corresponding group on the T-dual side. 

The T-dual orientifold theory is IIA/$\Omega\cdot\CI_m$ if $m$ is
odd, and IIB/$\Omega\cdot\CI_m$ if $m$ is even, with $\CI_m$ the
involution (times appropriate factors of $(-1)^{F_L}$) that
reflects $m$ compact dimensions.  D-brane charges in these theories
will therefore be classified by the relative KR-group 
$\KR^{-n}(X\times\T^m,\,\T^m)$ (for some $n$), where the involution of 
KR-theory acts trivially on $X$, and reflects all the dimensions of $\T^m$.  
An interesting subtlety arises when we try to determine the value
of $n$ that corresponds to $\T^m$.  In Type IA theory, i.e., for $m=1$,  
one could use the alternative string theory definition of $\KR^{-1}(X)$ 
to demonstrate that the appropriate value of $n$ is in fact $n=1$ \phk .  
{}For $m>1$, however, we do not seem to have that option.  Instead, we will 
proceed as follows.

Agreement with the known spectrum of {\it supersymmetric\/} D-branes 
determines that $n=m$ mod~4.  Since the Bott periodicity of
KO-theory is eight, this leaves an uncertainty as to whether $n$ equals $m$ 
or $m+4$.  We claim that the correct value is $n=m$, and the D-brane charges 
in the Type I T-dual models are classified by the relative group
\eqn\eekrkrm{\KR^{-m}(X\times\T^m,\,\T^m).}
Indeed, iterating \eecalckr , and using the $(1,1)$ Bott periodicity of 
KR-theory, one can show that 
\eqn\eekrcalcp{\KR^{-n}(X\times\T^m,\T^m)=\bigoplus_{p=0}^m\pmatrix{m\cr
p\cr}\tilde\KR{}^{n,p}(X)=\bigoplus_{p=0}^m\pmatrix{m\cr p\cr}
\tilde\KO{}^{p-n}(X);}
this coincides with \eetorko\ for $n=m$.%
\foot{Heuristically, the exponent $-m$ in the relevant KR-group has to do with
the fact that $\S^1$ with the reflection $\CI$ behaves effectively
as a ``sphere of dimension minus one'' in KR-theory: while the wedge product
with $\S^1$ with
trivial involution lowers the exponent of the KR-group by one, the wedge
product with $\S^1$ with the reflection raises the exponent by one; cf.\ the
appendix.}
Thus, we again get a T-duality isomorphism
\eqn\eeclassifvectdual{\KR^{-m}(X\times\T^m,\T^m)\cong\KO(X\times\T^m,\T^m).}
The corresponding spectrum of both the BPS and the ($\Z_2$) stable non-BPS 
D-branes is in precise agreement with the degeneracies of various wrapped 
branes.  Still, the precise bookkeeping of D-brane charges will involve 
various higher-dimensional analogs of the brane transfer operation studied 
above.  As a particularly interesting example, consider the classification of 
D-instantons in Type IIB on $T^2/\Omega\cdot\CI_2$, in terms of 
$\KR^{-2}(\S^8\times\T^2,\,\T^2)=\Z\oplus\Z_2\oplus\Z_2\oplus\Z_2$.  This 
group contains fewer charges than naively expected; this is again 
resolved by brane transfer operations, which now involve objects that wrap 
zero, one, or both dimensions of the compact $\T^2$.  We leave details as an 
exercise to the interested reader.  

\newsec{Type $\TI$ Theories}

\noindent
When considering theories in nine dimensions, there exists a
natural extension of Type IA which involves replacing one of its
two $\O8^-$ planes with an $\O8^+$ plane \gp .  We will refer to it as
Type $\TIA$ theory.  This theory requires no D8-branes, and so has
no gauge group, yet it still contains interesting non-BPS stable D-branes 
in its spectrum.  We will demonstrate that some charges which are locally 
stable at one type of orientifold plane will now become unstable due to the 
possibility of moving to another type of orientifold plane and annihilating 
there.  Also, stretched $\Z_2$ charged D-branes will no longer be connected 
with the transfer of stuck branes from one plane to the other.

To classify the possible D-brane charges of Type $\TIA$ we
will once again analyze possible tachyon backgrounds of unstable
D9-branes using K-theory.  For this purpose, it is easier to start
with the T-dual of Type $\TIA$, Type $\TI$ theory.  This T-dual
was worked out in \ewvect , and consists of gauging a
$\Z_2$ symmetry of IIB on a circle which is realized by 
composing worldsheet parity reversal ($\Omega$) with a
half-circumference shift along the circle.  The natural K-group of D-brane
charges for Type $\TI$ is then the relative group
$\KR(X\times\S^{0,2},\,\S^{0,2})$.%
\foot{We are again using the $\S^{p,q}$ notation reviewed in the appendix.}
Since $\S^{0,2}$ is just a circle in $\R^2$ with both dimensions reflected, 
the involution of KR-theory indeed acts on $\S^{0,2}$ by the required shift.  

The Type $\TI$ K-group, $\KR(X\times\S^{0,2})$, is known
in the mathematics literature to be isomorphic to the K-group
$\KSC(X)$ of self-conjugate bundles on $X$(see \refs{\mfareality,\psgreen,
\anderson} and the appendix). We bring this up because $\KSC(X)$ has
several nice features.  First of all, using the relation to KSC-theory, 
we can show for the relative groups that 
\eqn\eekscperfour{\KR(X\times\S^{0,2},\,\S^{0,2})\cong\KR^{-4}(X\times\S^{0,2},
\,\S^{0,2}).}
The appearance of $\KR^{-4}(X\times\S^{0,2})$ is very interesting here, as 
this group associates a symplectic projection to $\Omega$.  Thus, 
the period of four indicated by \eekscperfour\ fits nicely with having both an
$\O8^-$ plane and $\O8^+$ plane in Type $\TIA$ -- indeed, it means that
orthogonal and symplectic groups appear on the same footing in this model.

Second, KSC groups have been calculated for all spheres $\S^n$ \psgreen , 
which means we can immediately read off the complete nine-dimensional
spectrum of Type $\TI$ and $\TIA$ D-branes:
\eqn\eetabthree{\vbox{\offinterlineskip \hrule
\halign{&\vrule#&\strut\ \ \hfil#\ \ \cr
height2pt&\omit&&\omit&&\omit&&\omit&&\omit&&\omit&&\omit&&\omit&&\omit&
&\omit&&\omit&\cr &D$p$-brane \ \
&&D8&&D7&&D6&&D5&&D4&&D3&&D2&&D1&&D0&&D$(-1)$&\cr
\noalign{\hrule}
height2pt&\omit&&\omit&&\omit&&\omit&&\omit&&\omit&&\omit&&\omit&&\omit&
&\omit&&\omit&\cr &$\tilde\KSC(\S^{8-p})$&&\Z\ &&$\Z_2$&&0 &&\Z\ &&\Z\
&&$\Z_2$&&0 &&\Z\ &&\Z\ &&$\Z_2\ \ $&\cr} \hrule}}
(Note that D8-branes appear on this list, even though the tadpole cancellation
argument will restrict the net number of D8-branes in the Type $\TIA$ vacuum
to zero.)

Unlike the relative K-theory groups that appeared in the previous
sections, $\KSC(X)$ does not naturally split into subgroups
related to wrapped and unwrapped branes.  Undeterred, we will try to 
analyze the physical spectrum listed above in terms of the wrapped 
and unwrapped D$p$-branes of Type $\TIA$ string theory, hoping to learn an 
interesting lesson when this strategy becomes inadequate.  

\subsec{Unwrapped D-branes of Type $\TIA$}

\noindent
   Our strategy for determining which nine-dimensional D$p$-branes come
from unwrapped D-branes of Type $\TIA$ (which are point-like along
the interval) is to use our knowledge of Type~IA theory to list
the stable D-brane spectrum near an $\O8^-$ plane, and to use a
simple period shift to list the stable D-brane spectrum near an
$\O8^+$ plane (cf.\ \gukov).  Modulo some identifications, this gives all the
D-branes in Table \eetabthree\ which come from unwrapped branes of
Type $\TIA$.  It will be instructive to follow this piecewise
analysis of the compactification manifold; its eventual inability
to explain the detailed spectrum of non-BPS states leads to
interesting conclusions.

In the vicinity of the $\O8^-$ plane, the D-brane spectrum is:
\eqn\eetabfour{\vbox{\offinterlineskip \hrule
\halign{&\vrule#&\strut\ \ \hfil#\ \ \cr
height2pt&\omit&&\omit&&\omit&&\omit&&\omit&&\omit&&\omit&&\omit&&\omit&
&\omit&&\omit&\cr &D$p$-brane
&&D8&&D7&&D6&&D5&&D4&&D3&&D2&&D1&&D0&&D$(-1)$&\cr \noalign{\hrule}
height2pt&\omit&&\omit&&\omit&&\omit&&\omit&&\omit&&\omit&&\omit&&\omit&
&\omit&&\omit&\cr &$\tilde\KO(\S^{8-p})$&&\Z\ &&$\Z_2$&&$\Z_2$&&0 &&\Z\
&&0 &&0 &&0 &&\Z\ &&$\Z_2\ \ $&\cr} \hrule}}
The $\O8^+$ plane differs from $\O8^-$ by interchanging $SO$ and $Sp$ 
projections.  Due to Bott periodicity between KO- and KSp-theory, the switch 
from $\O8^-$ to $\O8^+$ corresponds to swapping D$p$-branes with 
D$(p+4)$-branes, and leads to the following spectrum near the $\O8^+$ plane:
\eqn\eetabfive{\vbox{\offinterlineskip \hrule
\halign{&\vrule#&\strut\ \ \hfil#\ \ \cr
height2pt&\omit&&\omit&&\omit&&\omit&&\omit&&\omit&&\omit&&\omit&&\omit&
&\omit&&\omit&\cr &D$p$-brane\ \
&&D8&&D7&&D6&&D5&&D4&&D3&&D2&&D1&&D0&&D$(-1)$&\cr \noalign{\hrule}
height2pt&\omit&&\omit&&\omit&&\omit&&\omit&&\omit&&\omit&&\omit&&\omit&
&\omit&&\omit&\cr &$\tilde\KSp(\S^{8-p})$&&\Z\ &&0 &&0 &&0 &&\Z\
&&$\Z_2$&&$\Z_2$&&0 &&\Z\ &&0\ \ \  &\cr} \hrule}}
Now we need only combine the last two tables and compare with
Table \eetabthree .

Looking first at the $\Z$--valued D-brane charges, we see that
we can correctly account for the BPS D0-branes and D4-branes of
Type $\TIA$.  The fact that these appear in both Table \eetabfour\
and Table \eetabfive\ reflects the fact that two half-D0-branes on
the $\O8^-$ plane can combine to make a single D0-brane in the
bulk which then becomes a D0-brane on the $\O8^+$ plane, and
similarly for the half-D4-branes on the $\O8^+$ plane.  Since
half-D-branes are now limited to living on only one of the $\O8$
planes, there is no need of extra $\Z_2$ charges for brane
transfer operations.

Having been successful with the BPS D-branes that carry conventional 
RR charges, we now turn to the $\Z_2$-charged non-BPS D-branes of Tables
\eetabfour\ and \eetabfive .  Here we run into an interesting
puzzle: while these tables correctly account for the
$\Z_2$-charged D$(-1)$-brane, D3-brane and D7-brane of Table
\eetabthree , they also predict a $\Z_2$-charged D2-brane and D6-brane, 
which are however absent in Table~\eetabthree . 
The resolution of this puzzle reveals an interesting new effect.  
Take, for example, the stable $\Z_2$-charged non-BPS D6-brane
identified in Table \eetabfour\ near the $\O8^-$ plane. It
consists of a D6-brane and its mirror D$\bar 6$-brane, where the
usual tachyon between the two has been removed by the orientifold
projection. Just like the Type I non-BPS D7-brane in section~3, this
system carries a $U(1)$ gauge group, and can separate in a
symmetric fashion and transfer over to the other $\O8$ plane. In
Type $\TIA$, however, the orientifold projection is different at
the other $\O8$ plane, and therefore the tachyon is no longer removed.  This 
implies that the non-BPS D6-brane locally stable near the $\O8^-$ plane is no
longer stable in the global theory.  

This effect has important consequences for the analysis of non-BPS 
D-branes on compact manifolds.  Typically, when looking at the BPS spectrum of
D-branes near singularities of a compact manifold such as K3 orientifolds, 
one can look piecewise at the singularities (i.e., approximate them
with ALE spaces) and add the corresponding spectra (being careful
to match the bulk D-branes).  We now see that for stable non-BPS
D-branes this is a dangerous procedure, as D-branes locally stable
near one kind of singularity can become unstable due to other
singularities in the complete space.

Now that we have successfully accounted for the unwrapped D-branes, 
we can move on to look at the charges in Table \eetabthree\ which come
from wrapped D-branes of Type $\TIA$.

\subsec{Wrapped D-Branes of Type $\TIA$}

\noindent
To examine how the wrapped D-branes of Type $\TIA$ contribute
to the nine-dimensional spectrum listed in Table \eetabthree , it
is more convenient to shift to the T-dual Type $\TI$ point of
view.  We can now use a simple construction to build their dual
Type $\TI$ D-branes, which are now unwrapped. The IIB $\Z_2$ symmetry we
gauged to get Type $\TI$ strings included a half-shift along a
circle. Requiring that we respect the $\Z_2$ symmetry means we
match each unwrapped D$p$-brane with another D$p$-brane at the
opposite position along the circle for $p=1$ or 5, and with
a D$\bar p$-brane for $p= -1$, 3, or 7.  The first
configuration is BPS, and correspondingly yields the stable
$\Z$-charged D1-brane and D5 brane in Table \eetabthree .  Note
that these D-branes will carry a $U(N)$ gauge group, and will
correspond to doubly wrapped BPS D2-branes and D6-branes in the
Type $\TIA$ theory.  

The second kind of configuration above is more
interesting, as it yields stable non-BPS D-branes.
It is clear that these states carry a $\Z_2$ charge,
since when two of them are present there is an
allowed motion which enables the D-branes and D-antibranes to
annihilate, as can be seen in Fig.~3. 
Consequently, there appear to be two sources of $\Z_2$ $p$-brane charge
in Type $\TI$ for $p=-1,3,7$;
the first is due to wrapped $(p+1)$-branes (the T-duals of the
unwrapped Type $\TIA$ D-branes discussed in the previous subsection), 
and the second is due to the above $p-\bar{p}$ combinations.
There is no contradiction with the fact that K-theory predicts only
a single $\Z_2$ \eetabthree , since the above states are stable in 
complementary
regions of moduli space (Fig.~4).
\fig{$\Z_2$ annihilation of D-branes in Type $\TI$}{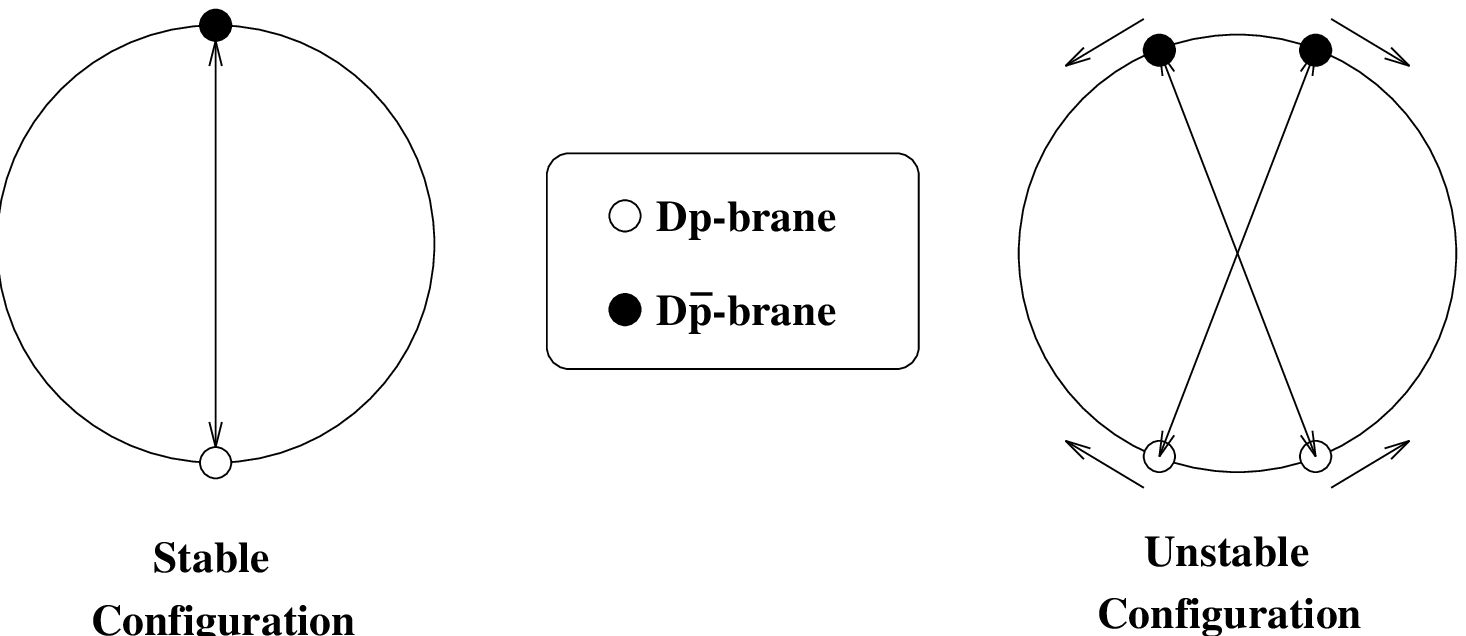}{5truein}

\fig{Instability of the D$p$-brane D$\bar p$-brane system}{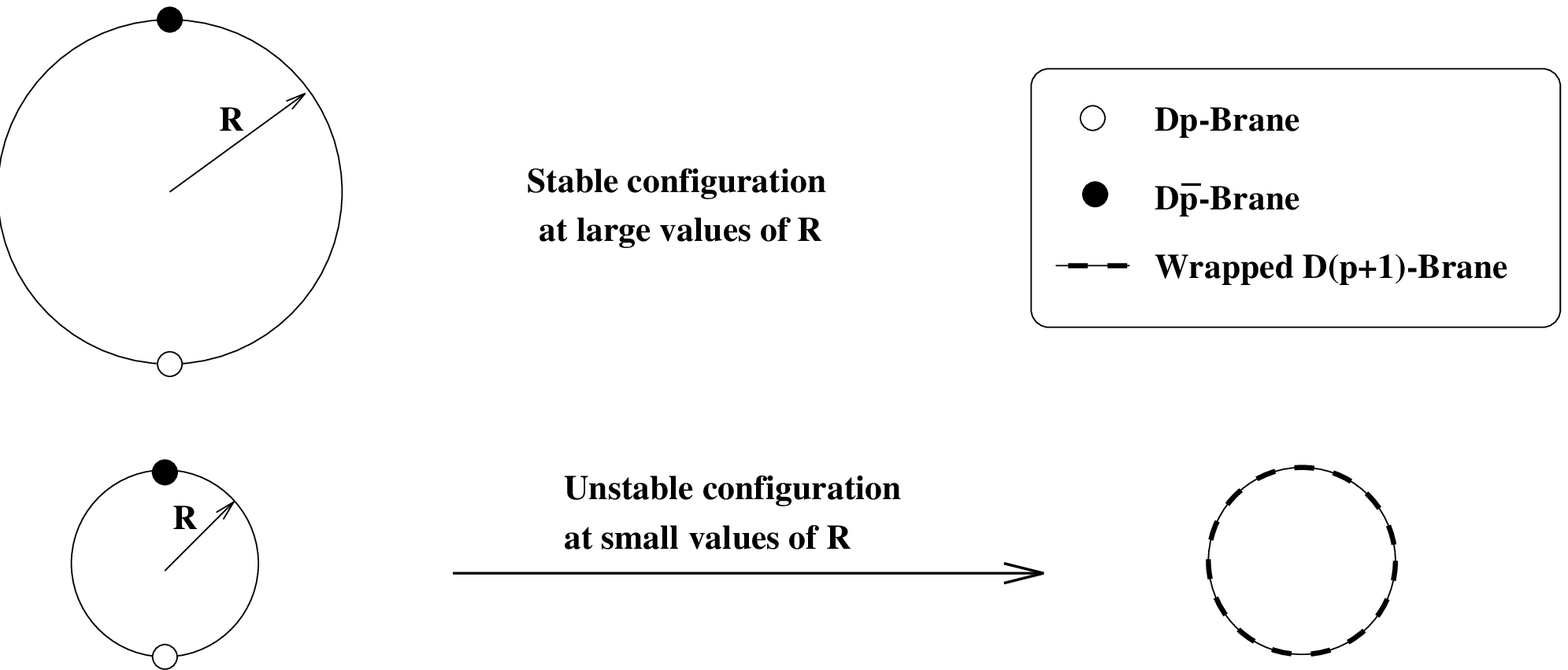}{6truein}

\newsec{Conclusions}

\noindent
In this paper, we used the picture of stable D-branes as topological defects 
in unstable brane systems to study the charges of 
stable non-BPS D-branes in string compactifications.  K-theory turns out to 
be a useful tool in this pursuit.  

We have seen that T-duality appears to be a manifest symmetry in K-theory.  
In Type II and Type I on $\T^m$, we have identified the relative K-theory 
groups on both sides of T-duality, and have demonstrated that they are 
isomorphic.  In the slightly more exotic Type $\TI$ Type $\TIA$ T-duality, we 
have identified the K-theory group of D-brane charges only on the Type $\TI$ 
side, and then demonstrated that it exactly corresponds to D-brane charges on 
the Type $\TIA$ side.  Of course, one should be able to identify the K-groups 
on {\it both\/} sides of T-duality from first principles, by studying how the 
orientifold group acts on unstable systems of spacetime-filling branes.  We 
certainly expect that such a direct analysis will confirm our findings, and 
will provide an extra check that T-duality is a manifest symmetry in 
K-theory.  It would also be instructive to extend our analysis of Type $\TI$ 
to all Type I models without vector structure \ewvect .  

In the process of identifying the D-branes which carry the charges predicted 
by K-theory, we have come across several interesting effects.  An apparent 
abundance of some $\Z_2$ charges in the string theory construction is 
resolved by brane transfer operations.  Other $\Z_2$ charges, apparently 
conserved locally near an orientifold plane, dissipate in the full theory 
due to the presence of another O-plane.  We believe that these phenomena 
occur in a more general class of compactifications, indicating that the 
piecewise analysis of stable non-BPS D-brane spectra should only be trusted 
when these phenomena are taken into account.  

T-duality is to be contrasted with other string theory dualities, such as 
S-duality of Type IIB string theory; whether or not there is an extension of 
K-theory that incorporates Type IIB S-duality -- and in particular explains 
NS states -- remains one of the many intriguing open questions of this 
framework.  (For a possible step in this direction, see \piljin .)  

\bigskip\medskip
\noindent We would like to thank Joe Minahan, Jan Nekov\'a\v r, 
Rahul Pandharipande, and Edward Witten
for useful conversations.
This work has been supported by DOE Grant DE-FG03-92-ER~40701.
P.H. is also supported by a Sherman Fairchild Prize Fellowship.
\vfill\break
\appendix{A}{Useful facts in K-theory}

In this somewhat extensive appendix, we present a summary of some basic 
notions of K-theory (and its connection to string theory), as well as 
technical details of various K-theory calculations needed for some of our 
arguments in the body of the paper.  For a more comprehensive introduction to 
K-theory, the reader should consult
\refs{\karoubi,\spingeo,\husemoller,\atiyah}.  Some elementary K-theory facts 
that arise in the string theory context can also be found in \refs{\ewk,\phk}.

\bigskip\noindent
{\it Unitary K-theory $\K^{-n}(X)$}
\medskip

The unitary K-theory group $\K(X)$ is defined, for a given compact manifold
$X$,%
\foot{Unless explicitly stated otherwise, $X$ in this paper is always a
compact connected manifold.  K-theory can be extended to non-compact 
manifolds, as K-theory with compact support; one essentially defines $\K(Z)=
\tilde\K(\tilde Z)$, where $Z$ is the one-point compactification of $Z$.  
Thus, K-theory with compact support can be related to K-theory of compact 
manifolds, and we will not use it explicitly in this paper.}
as the group of equivalence classes of pairs of unitary bundles $(E,F)$
on $X$, where two pairs are declared equivalent if they can be made isomorphic
to each other by adding pairs of isomorphic bundles $(H,H)$.

The reduced group $\tilde\K(X)$ is defined as the kernel of the natural map
$\K(X)\rightarrow\K(x_0)$ to the K-theory group $\K(x_0)=\Z$ of a point $x_0$
in $X$, induced from the map $x_0\rightarrow X$.  The full K-theory group
splits canonically as $\K(X)=\tilde\K(X)\oplus\Z$.  In Type IIB string theory
on $X$, $E$ and $F$ are the Chan-Paton bundles of branes and anti-branes
wrapping $X$, and $\tilde\K(X)$ classifies invariant charges that can be
carried by general tadpole-cancelling configurations of brane-antibrane pairs,
modulo pair creation and annihilation.

Higher (reduced) K-theory groups $\tilde\K^{-n}(X)$ are defined by
\eqn\aeetwo{\tilde\K{}^{-n}(X)=\tilde\K(X\wedge\S^n).}
Here the wedge product is defined for two manifolds $X$ and $Y$ with
a marked point $x_0$ in $X$ and $y_0$ in $Y$, as the (topological) coset%
\foot{Given a closed submanifold $Y$ in a compact manifold $X$, the 
topological coset $X/Y$ is defined by shrinking $Y$ -- as a subset in $X$ -- 
into a point.}
\eqn\aeethree{X\wedge Y=X\times Y/(X\times y_0)\cup(Y\times x_0).}
(For example, $\S^1\wedge\S^1=\S^2$, $\S^n\wedge\S^1=\S^{n+1}$.)
The higher reduced groups are related to the unreduced groups by
$\K^{-n}(X)=\tilde\K{}^{-n}(X)\oplus\K^{-n}(x_0)$, where the K-theory
groups of a point are given by
\eqn\eepoint{\eqalign{\K^{-2p}(x_0)&=\Z,\cr
\K^{-2p-1}(x_0)&=0.\cr}}
Therefore $\K(X) = \tilde{\K}(X) \oplus \Z$ and 
$\K^{-1}(X) = \tilde\K{}^{-1}(X)$.

The definition of the higher K-theory group $\K^{-1}(X)$ (which classifies
Type IIA theory D-branes) in terms of the $p+1$-dimensional extension 
$X\wedge\S^1$ of the $p$-dimensional manifold $X$ is rather awkward for 
string theory purposes, as it invokes an extra spacetime dimension
$\S^1$.  There is an alternative definition of $\K^{-1}(X)$, in terms of
pairs $(E,\alpha)$ where $E$ is a bundle on $X$ and $\alpha$ is an
automorphism on $E$ (see, e.g., \karoubi, II.3.3).  It is, in fact, this
alternative definition that appears directly from the worldvolume dynamics of
spacetime-filling unstable D9-branes in Type IIA theory \phk .

K-theory is intimately related to homotopy theory.  K-theory groups of any
compact manifold $X$ can be understood as the groups of homotopy classes of
maps from $X$ to certain classifying spaces,
\eqn\aeehmtclsk{\eqalign{\tilde\K(X)&=[X,BU],\cr
\K^{-1}(X)&=[X,U].\cr}}
Here $U$ and $BU$ are the $N\rightarrow\infty$ limits of the unitary group 
$U(N)$ and the (group) coset $U(2N)/U(N)\times U(N)$, respectively.%
\foot{More generally, $\tilde\K{}^{-n}(X)=[X,\Omega^nBU]$, where $\Omega^nY$
is the $n$-th iterated loop space of $Y$.  One can prove that $\Omega BU$ is
homotopically equivalent to $U$, and $\Omega^2 BU$ is homotopically equivalent
to $BU$.  In conjunction with \aeehmtclsk , this fact leads to Bott
periodicity, $\K^{-n-2}(X)=K^{-n}(X)$.}

Given a closed submanifold $Y$ in a compact manifold $X$, one defines the
relative K-theory group $\K(X,\,Y)$ as follows.  Just like in the definition
of $\K(X)$, we start with a pair of bundles $(E,F)$ on $X$.  In addition, we
choose a ``trivialization'' along $Y$, i.e., an isomorphism 
$\alpha:E|_Y\rightarrow F|_Y$ between the restrictions 
of $E$ and $F$ to the submanifold $Y$.  One defines a certain equivalence
relation on such triples $(E,F,\alpha)$, declaring two such triples
equivalent if they can be made isomorphic by creation or annihilation
of triples $(H,H,{\rm id}_H)$ (see \karoubi , II.2.29 for the mathematicians'
definition).  Similarly, higher relative K-groups are defined by
\eqn\aeekreldef{\K^{-n}(X,\,Y)=\K(X\times\B^n,X\times\S^{n-1}\cup Y
\times\B^n),}
where $\B^n$ is the unit ball in $\R^n$, and $\S^{n-1}$ is its boundary 
sphere.  The relative groups \aeekreldef\ represent a generalization of the 
reduced groups 
$\tilde\K{}^{-n}(X)$, since one can write $\tilde\K{}^{-n}(X)=\K^{-n}(X,x_0)$
with $x_0$ a point in $X$.  Also, relative groups are related to the reduced 
groups by $\K^{-n}(X,Y)=\tilde\K{}^{-n}(X/Y)$ whenever both sides of this 
equation make sense.    

The relative groups $\K^{-n}(X,\,Y)$ are important because they connect the
groups of $X$ and $Y$ via the exact sequence
\eqn\aeeexactk{\K^{-n-1}(Y)\rightarrow\K^{-n-1}(X)\rightarrow
\K^{-n}(X,\,Y)\rightarrow\K^{-n}(Y)\rightarrow\K^{-n}(X)}
(valid for any $n\geq 0$), which is reminiscent of similar exact sequences
from cohomology theory.  In fact, K-theory is a {\it generalized\/} cohomology
theory -- it satisfies all the axioms of cohomology theory {\it except\/} for
the dimension axiom.

In the case of relative K-theory groups $\K(W,\,Y)$ that appear
in this paper, the pairs $W$, $Y$ are of a very special type, with 
$W=X\times Y$ for some manifold $Y$.  {}For such pairs (or more generally, 
whenever $W$ is a ``retract'' of $Y$), \aeeexactk\ can be reduced to the 
following split exact sequence (cf.\ \karoubi, II.2.29),
\eqn\aeerelatsplit{0\rightarrow\K^{-n}(X\times Y,\,Y)\rightarrow
\K^{-n}(X\times Y)\rightarrow\K^{-n}(Y)\rightarrow 0,}
thus leading to
\eqn\aeerelprodk{\K^{-n}(X\times Y)=\K^{-n}(X\times Y,\,Y)\oplus\K^{-n}(Y).}
This formula allows one to evaluate the relative group
$\K^{-n}(X\times Y,\,Y)$ once $\K^{-n}(Y)$ and $\K^{-n}(X\times Y)$ are
found.

So far we have reduced the calculation of the relative K-theory group
$\K^{-n}(W\times Y,\,Y)$ to the calculation of the K-theory groups of
$W\times Y$.  The latter can be expressed through K-theory groups of
$X$ and $Y$ with the use of the following formula (\atiyah , 2.4.8),
\eqn\eeredprodk{\tilde\K{}^{-n}(X\times Y)=\tilde\K{}^{-n}(X\wedge Y)\oplus
\tilde\K{}^{-n}(X)\oplus\tilde\K{}^{-n}(Y).}

The case of our primary interest in Section~2 is $Y=\S^1$.   Using
\eeredprodk\ together with $\tilde\K{}^{-n}(X\wedge\S^1)=\tilde\K{}^{-n-1}(X)$
and Bott periodicity, we get
\eqn\aeefour{\eqalign{\tilde\K(X\times\S^1)&=\K^{-1}(X)\oplus\tilde\K(X),\cr
\K^{-1}(X\times\S^1)&=\tilde\K(X)\oplus\K^{-1}(X)\oplus\Z.\cr}}
This allows us to determine $\K(X\times\S^1,\,\S^1)$ and $\K^{-1}(X\times\S^1,
\,\S^1)$ using \aeerelprodk , leading to \eekunnk\ and \eekunnkone .  

Alternatively, we can calculate $\K^{-n}(X\times\S^1)$ in a manner that keeps
track of the multiplicative structure of the theory.
Define $\K^{\#}(X)=\K(X)\oplus\K^{-1}(X)$.  $\K^{\#}(X)$ is a graded ring,
with the obvious $\Z_2$ graded structure.  We have a K-theory analog of the
K\"unneth formula,
\eqn\eekunneth{\K^{\#}(X\times Y)=\K^{\#}(X)\otimes\K^{\#}(Y),}
which is valid if either $\K^{\#}(X)$ or $\K^{\#}(Y)$ is freely generated
(see, e.g., \karoubi, Proposition IV.3.24).
Since $\S^1$ has freely generated K-theory groups, we can set $Y=\S^1$.
(Notice that this strategy for calculating K-groups of $X\times\S^1$ would not
work in the case of KO-theory relevant for Type I, as the KO-groups of $\S^1$
are not freely generated.)  {}From \eekunneth\ we get
\eqn\eemultipk{\eqalign{\K(X\times\S^1)=(\K(X)\otimes\K(\S^1))\oplus
(\K^{-1}(X)\otimes\K^{-1}(\S^1)),\cr
\K^{-1}(X\times\S^1)=(\K(X)\otimes\K^{-1}(\S^1))\oplus
(\K^{-1}(X)\otimes\K(\S^1)).\cr}}
Using $\K(\S^1)=\Z$ and $\K^{-1}(\S^1)=\Z$, we again recover \aeefour , which 
is instrumental in our proof of T-duality between Type IIA and Type IIB 
theories in Section~2.  With the insight from \eemultipk , Type II T-duality 
can thus be traced back to the fact that K-theory groups of $\S^1$ have 
shortened periodicity, $\K^{-m}(\S^1)=\K^{-m-1}(\S^1)=\Z$.   Also, 
using \eemultipk , the fact that T-duality swaps wrapped and unwrapped branes 
corresponds to the fact that under the isomorphism \eetdualityii\ of the 
K-groups, $\tilde\K(X)\otimes\K(\S^1)$ maps to 
$\tilde\K(X)\otimes\K^{-1}(\S^1)$ (and similarly for $\K^{-1}(X)$), with 
$\K(\S^1)$ factors and $\K^{-1}(\S^1)$ factors interchanged.  

\bigskip\noindent
{\it Orthogonal K-theory $\KO^{-n}(X)$ and Symplectic K-theory $\KSp^{-n}(X)$}
\medskip

$\KO(X)$ is the group of virtual real bundles, defined by replacing complex
bundles with real bundles in the definition of $\K(X)$ groups.  Higher KO
groups are again defined via
\eqn\eehigherkodef{\tilde\KO{}^{-m}(X)=\tilde\K(X\wedge\S^m).}
Just like in the unitary case, the full KO-groups are related to the reduced
groups $\tilde\KO{}^{-m}(X)$
by
\eqn\eeredunredko{\KO^{-m}(X)=\tilde\KO{}^{-m}(X)\oplus\KO{}^{-m}(x_0),}
with $x_0$ a point in $X$.  The key to the appearance of KO-theory in the
bound-state construction of Type I D-branes is again its relation to homotopy
theory.  Just as in the unitary case, we have $\KO^{-n}(X)=[X,\Omega^nBO]$,%
\foot{More exactly, the classifying space of KO-theory is $BO\times\Z$, where 
the extra factor of $\Z$ is needed to explain $\KO({\rm pt})=\Z$ (see, e.g., 
\karoubi, II.1.34).}
where $BO$ is defined as the large-$N$ limit of $O(2N)/O(N)\times O(N)$, and
$\Omega BO\cong O$ can be similarly approximated by $O(N)$.  In this case, the 
statement of Bott periodicity $\KO^{-m}(X)=\KO^{-m-8}(X)$ follows from the 
fact that $\Omega^{m+8}BO$ is homotopically equivalent to $\Omega^mBO$.

By replacing $O(N)$ with $Sp(N)$, and real bundles with symplectic bundles,
one can similarly define the symplectic K-theory groups $\KSp^{-n}(X)$.
Bott periodicity can be refined to show that $\KO^{-n}(X)=\KSp^{-n-4}(X)$ for
any $n$, which means that any calculation in KSp-theory can be done in
KO-theory; therefore, we will not discuss KSp-theory separately in this
appendix.

Relative K-theory groups $\KO^{-n}(Z,\,Y)$ are defined by replacing complex 
bundles with real bundles in the definition of $\K^{-n}(Z,\,Y)$ reviewed 
above.  
{}For our purposes, we will again be interested in relative groups for a 
special class of pairs, $\KO^{-n}(X\times Y,\,Y)$; for such pairs, one can 
relate the relative group to $\KO^{-n}(X\times Y)$ and $\KO^{-n}(Y)$ via the 
following split exact sequence,
\eqn\aeesplitko{0\rightarrow\KO^{-n}(X\times Y,\,Y)\rightarrow\KO^{-n}(X\times
Y)\rightarrow\KO^{-n}(Y)\rightarrow 0,}
leading to
\eqn\aeerelprodko{\KO^{-n}(X\times Y)=\KO^{-n}(X\times Y,\,Y)\oplus
\KO^{-n}(Y).}

The basic formula for calculating KO groups of products is again
\eqn\eeredprodko{\tilde\KO(X\times Y)=\tilde\KO(X\wedge Y)\oplus\tilde\KO(X)
\oplus\tilde\KO(Y).}
In the case of our main interest, $Y=\S^1$, we obtain (using
$\tilde\KO(\S^1)=\Z_2$, and $\tilde\KO(X\wedge\S^1)=\tilde\KO{}^{-1}(X)$)
\eqn\eeredko{\tilde\KO(X\times\S^1)=\tilde\KO^{-1}(X)\oplus\tilde\KO(X)
\oplus\Z_2.}
This formula is used in Section~3 on the Type I side of the proof of T-duality
between D-brane charges.

\bigskip\noindent
{\it Real K-theory $\KR^{p,q}(X)$}
\medskip

KR-theory (introduced, under the name of ``Real K-theory,'' by Atiyah
\mfareality) is a generalized theory that includes unitary K-theory, KO-theory
and KSp-theory (as well as the ``self-conjugate'' KSC-theory to be discussed
below) as special cases.  The corresponding groups $\KR(X)$ are defined
for $X$ a manifold with a selected involution $\tau$.  Basic objects are now
pairs of bundles $(E,F)$ with antilinear involution on both $E$ and $F$ that
commutes with $\tau$ on $X$.  Thus, $\KR(X)$ would be the group of virtual
bundles with involutions on $X$.

Just like in KO-theory, one can define higher groups $\KR^{-m}(X)$, by
\eqn\eehigherkr{\tilde\KR{}^{-m}(X)=\tilde\KR(X\wedge\S^m);}
the involution $\tau$ of $X$ is extended to the involution of $X\wedge\S^m$
that acts trivially on $\S^m$.

More generally, we can consider replacing $\S^m$ in \eehigherkr\ by spheres
with non-trivial actions of the involution.  Consider
$\R^{p,q}$, as a real manifold of dimension $p+q$ with coordinates
$(x^1,\ldots,x^p,y^1,\ldots,y^q)$, and with involution that takes
$(x,y)\rightarrow(x,-y)$.  Similarly, one defines $\S^{p,q}$ as the unit 
sphere (of dimension $p+q-1$) in $\R^{p,q}$ with respect to the flat Euclidean 
metric.%
\foot{Notice that our convention for $\R^{p,q}$ and $\S^{p,q}$
coincides with that of Atiyah \mfareality , and is opposite to that of Karoubi
\karoubi .}

Now, we can define a two-parameter set $\KR^{p,q}(X)$ of higher KR-theory
groups, by
\eqn\eetwoparkr{\tilde\KR{}^{p,q}(X)=\tilde\KR(X\wedge\tilde\R{}^{p,q}),}
where $\tilde\R{}^{p,q}$ is the one-point compactification of $\R^{p,q}$ 
(i.e., $\tilde\R{}^{p,q}$ is topologically a $p+q$-sphere).  
By definition, \eehigherkr\ are related to \eetwoparkr\ by
$\KR^{-m}(X)=\KR^{m,0}(X)$.  One can define relative K-theory groups 
$\KR^{-n}(Z,\,Y)$, again by repeating steps used in the definition of relative 
K-groups in K-theory and KO-theory.  

Bott periodicity in KR-theory states that $\KR^{p,q}(X)=\KR^{p+1,q+1}(X)$, and
$\KR^{-m}(X)=\KR^{-m-8}(X)$.  Due to the first relation, $\KR^{p,q}(X)$
depends only on the difference $p-q$, and one has
$\KR^{p,q}(X)=\KR^{q-p}(X)$.  It is interesting to notice that in KR-theory,
spheres with antipodal involutions play the role of negative-dimensional
spheres.

KR-theory is a generalization of both K-theory and KO-theory.  {}For
any given $X$ with trivial involution, we have
\eqn\eekrgener{\eqalign{\K^{-m}(X)&=\KR^{-m}(X\times\S^{0,1}),\cr
\KO^{-m}(X)&=\KR^{-m}(X).\cr}}
In particular, one can derive Bott periodicity in $\K(X)$,  $\KO(X)$ and
$\KSp(X)$ from the periodicities of KR-theory.

Now we are equipped to calculate the relative group $\KR^{-1}(X\times
\S^{1,1},\,\S^{1,1})$ that classifies D-brane charges in Type IA theory.
(The orientifold $\Z_2$ acts as a reflection on the circle $\S^{1,1}$, and
trivially on $X$.)  This relative group is again related to
$\KR(X\times\S^{1,1})$ via
\eqn\aeekrone{\KR^{-1}(X\times\S^{1,1})=\KR^{-1}(X\times\S^{1,1},\,\S^{1,1})
\oplus\KR^{-1}(\S^{1,1}).}
By repeating steps already familiar from the K and KO case, one obtains
\eqn\eecalckr{\eqalign{\tilde\KR{}^{-1}(X\times\S^{1,1})&=\tilde\KR{}^{-1}
(X\wedge\S^{1,1})\oplus\tilde\KR{}^{-1}(X)\oplus\Z\cr
&{}=\tilde\KR{}^{1,1}(X)\oplus\tilde\KR{}^{-1}(X)\oplus\Z=\tilde\KO(X)\oplus
\tilde\KO{}^{-1}(X)\oplus\Z,\cr}}
where we have used the $(1,1)$ periodicity of KR-theory, and the fact that
$\KR^\ast(X)=\KO^\ast(X)$ for the trivial orientifold action on $X$.  
Since $\tilde\KR^{-1}(\S^{1,1})=\KR^{-1+1}({\rm pt})=\KO({\rm pt})=\Z$, our 
result \eekrcalcp\ follows from \aeekrone\ and \eecalckr .  

\bigskip\noindent
{\it Self-Conjugate K-theory $\KSC^{-n}(X)$}
\medskip

Given a compact manifold $X$, one defines a self-conjugate bundle on $X$
as a bundle $E$ equipped with an antilinear automorphism $\beta:E\rightarrow
E$.  Self-conjugate K-theory $\KSC(X)$ (see \refs{\psgreen,\anderson,
\mfareality} and \karoubi\ III.7.13-15) is then defined by imposing a stable 
equivalence relation on self-conjugate bundles (i.e., on pairs $(E,\beta )$),
whereby two pairs are equivalent if their sums with a third self-conjugate 
bundle are isomorphic (as self-conjugate bundles).  Higher KSC groups are 
again defined via $\tilde\KSC{}^{-n}(X)=\tilde\KSC(X\wedge\S^n)$.  The 
classifying space $BSC$ of self-conjugate K-theory is described in \psgreen ; 
KSC-groups are then related to homotopy theory via $\KSC(X)=[X,BSC]$.  

One can prove that Bott periodicity of the self-conjugate K-theory is four.
This can be shown either by a direct analysis of the homotopy properties of
the classifying space \refs{\psgreen,\anderson} (and showing that it is
homotopically equivalent to its fourth loop space), or by proving a relation
between KSC-theory and KR-theory,
\eqn\aeeksc{\tilde\KSC(X)=\tilde\KR(X\times\S^{0,2}),}
and using Bott periodicity of KR-theory (see \mfareality\ for details).

The relation \aeeksc\ between KSC and KR groups plays a central role in our
analysis of T-duality between Type $\TI$ and Type $\TIA$ in Section~4.

\listrefs
\end